# On the Alternative Theories of Cosmology

A. Stepanian [1] [*], M. Kohandel [2] [†]

[1] Department of Physics, University of Tehran, Tehran, Iran

[2] Department of Physics, Alzahra University, Tehran, Iran

**Abstract**

In this article first we present a brief review of some alternative theories of cosmology. Then by referring to some of the main works done in these theories we show that how they can describe the phenomena which are explained nowadays in the framework of standard model of cosmology(SM).

## 1   Introduction

Nowadays there are many models of cosmology trying to describe the mechanism of our universe as a whole. The most famous and popular model is the Big Bang cosmology, often called the Standard Model(SM) , according to which the universe was created through a gigantic explosion in 14 billion years ago. As the SM's theoretical predictions are almost compatible with observational data, cosmologists has spent much of their time to improve and make it more precise. However, there are some who believe that the theory is wrong and a new theory should be used. They have developed different models and theories. But these theories have not become as popular as the SM. So they have too many problems which have not been discussed yet. In this article we speak briefly about four most famous theories, by dividing them into Machian and non-Machian theories. It is to be mentioned that we do not defend these theories nor refuse them. Our aim is to make interested audiences be more familiar with them.

---

[*] Email address: astepanian@ut.ac.ir
[†] Email address: kohandel.mahsa@gmail.com

# 2 Machian and non-Machian models

In 19-th century the German philosopher and scientist Ernst Mach published his ideas about the origin of inertia. He concluded that the inertial mass of any object depends on all other objects in the universe. His article influenced many physicists including Einstein, such that he tried to implement Mach's ideas in his General Theory of Relativity, which is the main theoretical framework of all modern cosmological theories. Regarding Mach's principle as an important factor for better understanding the universe, we divide the alternative theories into Machian and non-Machian models.

## 2.1 Machian theories

### 2.1.1 Quasi Steady State Cosmology (QSSC)

The quasi steady state cosmology is regarded as the serious competitor of SM. The main contributors were Hermann Bondi, Thomas Gold, Fred Hoyle, Geoffrey Burbidge and Jayant V. Narlikar. The theory is the evaluated form of the older steady state one.

In QSSC one begins with a principle named Perfect Cosmological Principle (PCP). This is the stronger version of Weyl principle. PCP states that our universe in the large scale remains unchanged with time, in addition to what states the Weyl principle.

By considering this principle, first of all we conclude that the Hubble parameter must be time independent, thus:

$$H = \frac{\dot{a}(t)}{a(t)} = cte \rightarrow a(t) = \exp(Ht)$$

Thus the space-time must be in this form:

$$ds^2 = c^2 dt^2 - e^{2Ht}[dr^2 + r^2 d\Omega^2] \quad [1]$$

## Action and equations [4]

As mentioned above the founders of the QSSC tried to implement the Mach's principle in the theory in one side and on the other side they wanted to construct the theory in such a way that it becomes conformal invariant. By regarding these two factors they replaced the modified Einstein-Hilbert action

$$\mathcal{A} = \frac{1}{16\pi G} \int (R + 2\Lambda)\sqrt{-g}\, d^4x$$

by

$$\mathcal{A} = \sum_a \int m_a(A)\, da$$

Where $m_a(A)$ indicates the object inertial mass of object **a** at the point A and the sum is over all objects in the universe.

This action is originated from the action that F.Hoyle and J.V Narlikar proposed for their steady state cosmology, where they claimed that for an object like **a** in the universe we have

$$m_a(A) = \lambda_a \sum_{b \neq a} m_b(A)$$

Which simply means that $m_a$ at the point A, can easily be calculated by the other object's mass, and $\lambda_a$ is the intrinsic coupling constant for **a**. It is easy to show that if we have no particle in the universe then $m_a(A) = 0$.

For this theory we can also define " advanced " and " retarded " solutions just like what we do for classical electrodynamics, but in this case we have to replace the " Mass field " M(x) wih the ordinary "Electromagnetic potential " . Thus this Mass field must satisfy

$$\Box M(x) + \frac{1}{6} R\, M(x) = \sum_a \int \frac{\delta_4(x, A)}{\sqrt{-g(A)}}$$

This is the wave equation for M(x). This equation contains both advanced and retarded solutions. By removing the ambiguities for these solutions and putting

$$m_a(A) = M(A) \quad , \quad m_b(B) = M(B)$$

Our equation for proposed action becomes

$$(R_{\mu\nu} - \frac{1}{2}g_{\mu\nu}R) = \frac{16}{M^2}[-T_{\mu\nu} + M_\mu M_\nu - \frac{1}{2}g_{\mu\nu}g^{mn}M_m M_n + g_{\mu\nu}\Box\frac{1}{6}M^2 - M_{;\mu\nu}]$$

Under conformal transformations M(x) transforms as

$$M(x)\Omega^{-1}(x) = M'(x)$$

Thus by choosing a conformal frame where $M'(x)$ is a constant, say $M_0$, our equation become

$$R_{\mu\nu} - \frac{1}{2}g_{\mu\nu}R = \frac{6}{M_0^2}T_{\mu\nu}$$

Now by taking $G = \frac{3}{4\pi M_0^2}$

$$R_{\mu\nu} - \frac{1}{2}g_{\mu\nu}R = -8\pi G T_{\mu\nu}$$

which is Einstein's field equation.

One of the main advantages of this theory is that the Newtonian Gravitational constant G has been achieved by the calculation of the theory rather than put it by hand. Another important fact is that by looking to achieved formula for G, one can find some similarities between G and the Planck mass. To be more precise

$$G = \frac{3}{4\pi M_0^2} \quad \text{and} \quad G = \frac{3\hbar c}{4\pi m_p^2}$$

Thus it seems that the Planck mass and the Planck particles play an important role in this new cosmology and they are not simply achieved by dimensional analysis.

## Matter Creation [1][2][4]

Now we start to explain the main difference between QSSC and SM. That is the creation of matter. In SM we have no reason for creation of matter. Big Bang theorists attribute this process to the gigantic explosion at the moment of creation. But in the QSSC since the universe is infinite and eternal, the creation can possibly take place at different times in different places.

In this scenario, the above introduced Planck particles play the main role. In QSSC there is $\mathcal{C} - field$ which is responsible for the matter creation process. The creation process starts when the energy of $\mathcal{C} - field$ equals to the $m_p^2$ (c=1). This process takes place almost in the vicinity of a strong gravitational field, for example in the presence of Schwarzschild black-hole when

$$\mathcal{C}^m \mathcal{C}_m \sim \frac{constant}{[1-\frac{2GM}{r}]}.$$

A Planck particle named **a** exists from $A_0$ to $A_0 + \delta A_0$, decays into n particles where n approximately is $6 \times 10^{18}$, where their mass field is $m^{(a_r)}(x)$, and satisfy the wave equation

$$\Box m^{(a_r)} + \frac{1}{6} R m^{(a_r)} + n^2 m^{(a_r)^3} = \frac{1}{n} \int_{A_0}^{A_0+\delta A_0} \frac{\delta_4(x, A)}{\sqrt{-g(A)}} da$$

The Planck particle **a** contributes $C^a(x)$, which also satisfies

$$\Box C^a + \frac{1}{6} R C^a + C^{a^3} = \int_{A_0}^{A_0+\delta A_0} \frac{\delta_4(x, A)}{\sqrt{-g(A)}} da$$

Thus we have a total C(x), where

$$C(x) = \sum_a C^a(x)$$

And m(x),

$$m(x) = \sum_a \sum_{r=1}^{n} m^{a_r}(x)$$

And the total mass

$$M(x) = m(x) + C(x)$$

By this our main gravitational equation becomes:

$$R_{\mu\nu} - \frac{1}{2} g_{\mu\nu} R = \frac{6}{m^2} [-T_{\mu\nu} + \frac{1}{6}(g_{\mu\nu} m^2 - m^2_{;\mu\nu}) + \left(m_\mu m_\nu - \frac{1}{2} g_{\mu\nu} m_i m^i\right) \\ + \frac{2}{3}\left(C_\mu C_\nu - \frac{1}{4} g_{\mu\nu} C_i C^i\right)]$$

This equation is again conformal invariant, so by changing this conformal frame to another one, one can finds a frame where $m(x)$ is constant,

$$m_0 \Omega(x) = m(x)$$

So just like what happened for action without $\mathcal{C} - field$,

$$8\pi G = \frac{6}{m_0^2}$$

$$R_{\mu\nu} - \frac{1}{2}g_{\mu\nu}R = 8\pi G\left[T_{\mu\nu} - \frac{2}{3}\left(C_\mu C_\nu - \frac{1}{4}g_{ik}C_i C^i\right)\right]$$

So the energy momentum tensor for $\mathcal{C}-field$ can be defined as

$$\mathcal{T}_{\mu\nu} = -\frac{2}{3}\left[C_\mu C_\nu - \frac{1}{4}g_{\mu\nu}C_i C^i\right].$$

In this conformal frame the action will be

$$\mathcal{A} = \frac{c^3}{16\pi G}\int R\sqrt{-g}\, d^4x - \sum_a m_a C \int ds_a - \frac{1}{20}\left(\frac{2}{3}\right)\int C_\mu C^\mu \sqrt{-g}\, d^4x + \sum_a \int C_\mu da^\mu.$$

The important feature of this action is that by taking the variation one can find

$$m_a c \frac{da^\mu}{ds_a}g_{\mu\nu} - C_\nu = 0$$

which is the overall energy-momentum conservation.

Since the coupling constant of $\mathcal{C}-field$ 's stress tensor is negative, it is easy to conclude that the created matter is thrown out after the creation. These processes are called mini-bangs. Mini-bangs, contrarily to Big Bang, are not singular.

As mentioned above according to PCP in a complete steady state universe, the space-time will be de Sitter. But since in QSSC we have mini-bangs the metric will be oscillatory.

## Some observational remarks [3]

Although in both SM and QSSC the chemical elements are synthesized, in QSSC we have no beginning for the universe and the creation take place in the nuclei of galaxies.

It is very striking that according to QSSC if the $He^4$ was synthesized from H in stars, the released energy after thermalization will have a black-body radiation, which is very close to CMB.

### 2.1.2 Fractal Cosmology

In mathematics and geometry fractals are objects which show a self-similarity in all degrees of resolution and all scales. We have also a branch in mathematics named fractal geometry. Mathematicians studied fractals for long time, but physicists did not show too much interest in it. But in 1980 by some works of Laurant Nottale and Yurij Baryshev, physicists became interested in it. It was shown that it is possible to attribute the self-similarity properties to our whole universe, by considering it as a fractal. Thus a new cosmological theory was born. In this sense for understanding the fractality of the universe a new theory, named Scale-Relativity constructed. So a new aspect of geometry entered to physics.

To implement the scale relativity one has to consider, as mentioned by Laurant Nottale, the following program.

i. Scale laws
ii. Induced effects
iii. Scale-Motion couplings

In the Scale-Relativity, space-time is considered as a fractal and then in ii and iii levels the concept of quantum physics and gauged fields are recovered.
This approach is very similar to what Einstein did, indeed he showed that the gravity can be manifested as the curvature in Riemannian geometry , by taking the same path and entering the new concepts , it is shown that quantum mechanics is the fractality of space-time. [9]

## Definition of fractals and some of their properties [9]

For every fractal we define two dimensions. One is the topological, $D_T$, and other is its fractal dimension, $D_F$.

Topological dimensions are always integers, for example for fractal dusts, curves and surfaces their topological dimensions are 0, 1 and 2 respectively.

Fractal dimensions are shown by $\frac{\log p}{\log q}$ where p is the number of its segments while q is the length of each segment.

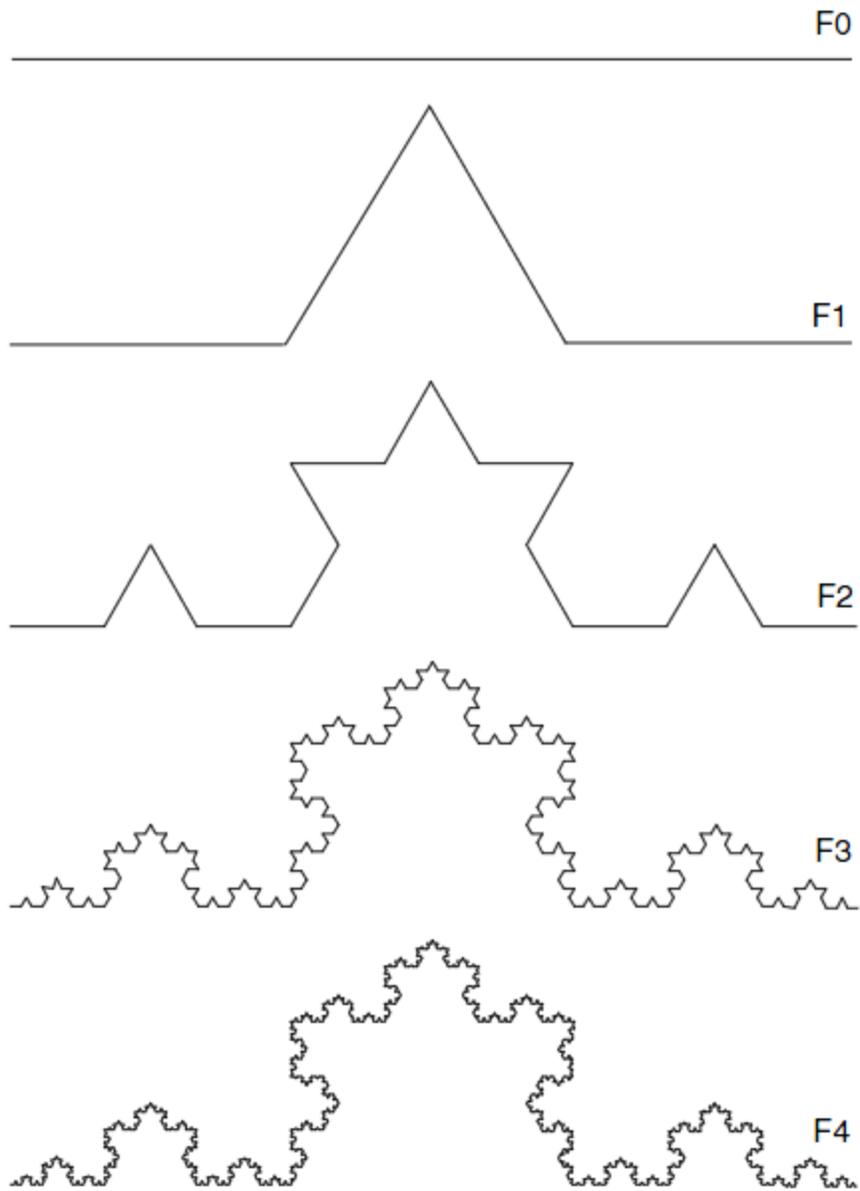

**Figure 1-Van Koch curve**

According to this definition, we are able to define the length of fractal, for example for a fractal curve:
$$\mathcal{L}(s,\epsilon) = s\left(\frac{\lambda}{\epsilon}\right)^{\tau} = \mathcal{L}(s,\lambda)\left(\frac{\lambda}{\epsilon}\right)^{\tau}$$

where **s** is some parameter and $\mathcal{E}$ called resolution. $\tau = D_F - D_T$, and $\mathcal{L}(s,\mathcal{E})$ is a renormalized curvilinear coordinate. This definition can be generalized too. For example

$$\mathcal{L}(S,\epsilon_X,\epsilon_Y) = S\left\{\left(\frac{\lambda_X}{\epsilon_Y}\right)^{\tau} + \left(\frac{\epsilon_Y}{\lambda_X}\right)^{\tau}\right\}$$

Where $\epsilon_X$ and $\epsilon_Y$ are resolutions in X and Y directions.

## Scale Relativity [9]

The theory of scale relativity looks like the same as the theory of motion relativity. In this sense the theory states that just like in motion relativity, where the motion is not absolute, the scale of arbitrary state is not absolute but it is defined relatively to others.

Briefly speaking if we want to find the laws of scale transformations we can do these replacements

$$X \leftrightarrow \ln \mathcal{L}$$

$$v \leftrightarrow \ln \frac{\epsilon}{\epsilon'} = \ln \rho$$

$$t \leftrightarrow \tau$$

In this sense for example for Galilean scale relativity, one finds

$$\ln \frac{\mathcal{L}(\epsilon')}{\mathcal{L}} = \ln \frac{\mathcal{L}(\epsilon)}{\mathcal{L}} + \tau(\epsilon)\ln \frac{\epsilon'}{\epsilon}$$

$$\tau(\epsilon') = \tau(\epsilon)$$

which is very similar to

$$X' = X + vt$$

$$t' = t$$

Just like Galilean and special relativity where $v$ is relative, in the scale relativity, it is the ratio of resolution, $\ln \frac{\epsilon}{\epsilon'}$, which becomes relative. This similarity is also applicable to special scale relativity and in this sense its symmetry group will be Lorentz group.

## Invariant length Scale [9]

The theory of scale relativity also indicates that it must be a length scale in the nature which is invariant under dilations. It introduces the length $\Lambda$. Similarly one can find an invariant time $\Lambda/c$.

It is also clear that in nowadays physics we have discovered the Planck units. So we can associate $\Lambda$ to the Planck length and $\Lambda/c$ to the Planck time, below these scales space and time cannot be conceptualized.

## Scale relativity, Cosmology and Mach's principle [8]

Scale relativity states that our universe looks like a fractal. By this consideration it tries to answer the problems of cosmology. Like QSSC it emphasizes on Mach's principle, by dividing it to three levels.

1. The motion of a system in which the observer experiences the inertial forces is defined only relatively to other masses.
   By considering the principle of equivalence in general relativity, it can be concluded that there is no global inertial system. In this sense, if we try to implement the Mach's principle with the principle of equivalence, then the only solution will be the whole universe.
2. If the distant masses do define the inertial systems, they must also determine the amplitude of inertial forces. But this leads us to a contradiction. For example in Schwarzschild black hole, there is no mass in infinity. For resolving this paradox it can be supposed that the energy of an object with the whole universe cancels its self-energy of inertial origin, so we have

$$E_1 = mc^2 \quad \& \quad E_2 = \frac{GmM}{R} \quad \rightarrow \quad \frac{GM}{Rc^2} \approx 1$$

   This result is too strange which indicates that we are inside a black hole. It can be added that in the context of SM we have no reasonable explanation for the above result.

Also in this sense we can recover the

$$\Omega = \frac{8\pi G\rho}{3H^2} = 1$$

Simply by substituting

$$M = {}^4/_3 \pi r^3 \rho \quad and \quad r = \frac{c}{H}$$

and by requiring that for Schwarzschild black hole $\frac{2GM}{c^2 r} = 1$.

So we will have

$$2G \frac{4}{3} \pi \rho \frac{c^3/H^3}{C^3/H} = \frac{8\pi G \rho}{3H^2} = 1$$

3. In this level it can be possible to answer the question about the origin of the elementary particle's mass. As Mach concluded, just like the existence of relative motion, we can only speak about the relative mass i.e. there is no absolute mass. So we have

$$m_1 r_1 = m_2 r_2 \rightarrow \frac{m_1}{m_2} = \frac{r_2}{r_1}$$

By using $E = mc^2$ and $Gm_p^2 = \hbar c$, the gravitational force between two objects with masses $m_1$ and $m_2$ is

$$F = \hbar c \frac{\left(\frac{m_1}{m_p}\right)\left(\frac{m_2}{m_p}\right)}{r^2}$$

So the result can be stated in three different ways
- No preferential scale of mass do exists in nature.
- We have a preferential scale of mass and it is $m_p$.
- We have a preferential scale of mass called the scale of elementary particle mass scales.

For a nucleon mass denoted by $m_n$, it can be shown that

$$\left(\frac{M_u}{m_n}\right)^{\frac{1}{2}} \sim \left(\frac{m_p}{m_n}\right)^2 \sim \left(\frac{M_u}{m_p}\right)^{\frac{2}{3}}$$

where $M_u$ is the total mass of observable universe and $m_p$ is the Planck mass.

## Solving the Horizon problem [8]

By using the rules of scale relativity, the zero instant of SM is removed. It is replaced by the Planck's. Although this new time owns all the properties of t=0, everything becomes divergent, in this new formalism $t = \frac{\Lambda}{c}$ is not excluded from the evolution of the universe. By this the horizon problem can be solved with scale transforming and without any inflationary scenario. In scale relativity there is an inflation of light cone at $=\frac{\Lambda}{c}$, the Planck time. The fact that Λ is invariant under dilations means that when observed at resolution Λ, the distance between any two points reduce to Λ itself.

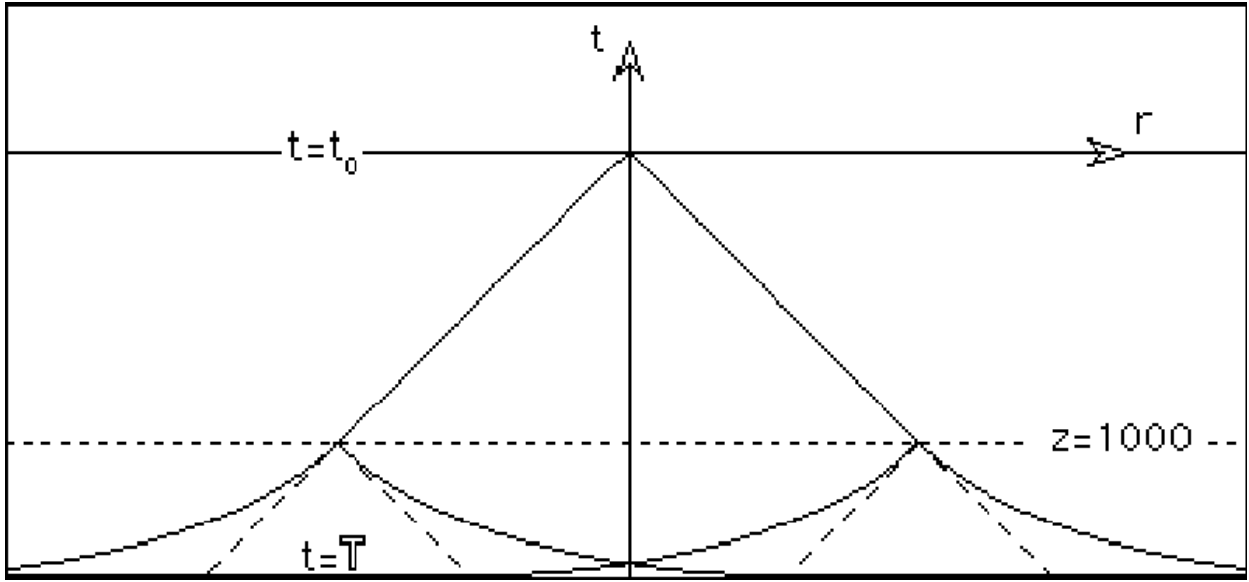

Figure 2-Inflation of light cone at the Planck time

## Large number coincidence [11]

As mentioned above, scale relativity states that our universe looks like a fractal and has the self-similarity feature. This means that the universe as a whole is too similar to elementary particles. This statement becomes more interesting when we compare the lifetime, radius and mass of whole universe with an elementary particle's ones.

Then we see

$$\frac{T}{t} \sim \frac{R}{r} \sim \left(\frac{M}{m}\right)^{\frac{1}{2}}$$

From here it is also possible to re-scale the Planck's constant $h$ by the same rate and obtained a new constant H, by which one will be able to describe the Gödel's spin and Kerr limit.

One of the strange and mysterious relations in theoretical physics is named Large Number Coincidence

$$\hbar^2 H_0 = G m_n^3 c$$

It seems that this coincidence relates all fundamental constants in physics together. It can be shown that this result can simply achieved by applying 2$^{nd}$ level of Mach's principle and the mass relation between elementary particles, universe and the Planck particle. So

$$mc^2 = \frac{GmM_u}{R_u}$$

$$\frac{c^2 R_u}{G} = \frac{m_p^4}{m_n^3} = \frac{\hbar^2 c^2}{G^2 m_n^3}$$

$$\frac{R_u}{G} = \frac{\hbar^2}{G^2 m_n^3}$$

$$G m_n^3 = \frac{\hbar^2}{R_u}$$

$$G m_n^3 = \frac{\hbar^2}{c/H_0}$$

$$G m_n^3 c = \hbar^2 H_0$$

## Strong Gravity [10]

One of the interesting aspects of Scale-Relativity is that in its context one is able to derive the concept of strong gravity. Strong gravity, by comparing the size of cosmos and elementary particles deduce that the strong force for elementary particles, especially hadrons, is similar to gravity for cosmos

$$\frac{T}{t} = \frac{R_u}{r_u} \sim \frac{S_s}{S_g}$$

Where $S_s$ and $S_g$ are the strength of Strong force and Gravity respectively. Thus it can be concluded that the cosmos and hadrons are systems with scales N and N-1 respectively, whose fractal dimension is D=2.

# Confinement and Asymptotic Freedom [10]

According to this self-similarity, it is natural to try to explain the physics of a hadron in terms of " Re-scaled Gravity ". For this, one must define the strength of gravity and strong force

$$S_g = \frac{Gm^2}{\hbar c} \quad \& \quad S_g = \frac{Ng^2}{\hbar c}$$

where N and g are the strong force constant and the color charge respectively. So the equation for a hadron becomes

$$R_{\mu\nu} - 1/2\, g_{\mu\nu} R - \Lambda g_{\mu\nu} = -KNT_{\mu\nu}$$

$$K = \frac{8\pi}{c^4}$$

In the case of a spherically symmetric distribution of $g'$, calculations show that a parton will feel a force like :

$$F = g'' \frac{d^2 r}{dt^2} = -1/2\, c^2(g'') \left(1 - \frac{2Ng'}{c^2 r} + 1/3\, \lambda r^2\right)\left(\frac{2Ng'}{c^2 r^2} + \frac{2}{3}\lambda r\right)$$

At very large distance this force simply reduced to

$$F \approx -g''c^2 \frac{\lambda r}{3}$$

which indicates a strong attractive force. This is the so-called Confinement.

On the other side for $r \geq 1$ fm we can find

$$F \approx -1/3\, g''c^2 \lambda (1 + \frac{\lambda r^3}{3} - \frac{Ng'}{c^2})$$

The effective potential will be

$$V_{eff} = \frac{1}{2} g''c^2 [2(\frac{Ng'}{c^2})^2 \frac{1}{r^2} - 2\frac{Ng}{c^2}\frac{1}{r} - 2\frac{Ng'}{3c^2}r + \frac{1}{2}\left(\frac{\lambda}{3}\right)^2 r^4] + \frac{\left(\frac{J}{g''}\right)^2}{2r^2}.$$

When the " re-scaled general relativity " becomes "Newtonian strong force "

The effective potential becomes

$$V_{eff} \approx \frac{-Ng''g}{r} + \frac{\left(\frac{J}{g''}\right)^2}{2r^2}.$$

When

$$r_e = \frac{J^2}{Ng'g''}$$

The effective potential vanishes and we have an asymptotic freedom.

## Hadrons as micro black holes [12]

By continuing the same path, it is also possible to get another interesting similarity, this time between elementary particles and black holes. It is seen that Hadrons and Kerr-Newman black holes are similar to each other in the sense that:

1. Both of them have a gyromagnetic ratio 2.
2. None of them have electric dipole moment
3. Both can be recognized by their spin, mass and charge
4. In both Kerr-Newman black holes and hadrons we have a relation between their mass and spin $J \propto M^2$

## Mass of protons and α-particle's radius

By re-scaling the gravitational force to the strong force some calculations can be done. For example in black hole physics we have

$$J = aG_0 \frac{M^2}{c}$$

Where **a** is a dimensionless parameter for spin. For Kerr-Newman black holes by re-scaling this relations we have

$$j_p = \frac{1}{4} N \frac{m^2}{c} \quad \text{(for a proton)}$$

where for proton, the spin parameter becomes ¼ which is half of its total spin integer.

Then

$$j_{proton} \approx \frac{\hbar}{2} = \left(\frac{1}{4}\right) N \, m^2/c$$

$$m = \left(\frac{hcN}{\pi}\right)^2$$

$$m = 1.7 \times 10^{-24} g$$

Surprisingly this result is in agreement with experiment since the experimental result is $1.67 \times 10^{-24}$

To calculate the radius of α-particle, a particle with spin zero, it must be regard as Schwarzschild black hole.

Since the radius of Schwarzschild black hole is

$$R = \frac{2GM}{c^2}$$

So for α-particle it becomes

$$r_\alpha = \frac{2Nm_\alpha}{c}$$

$$r_\alpha = 3.26 \times 10^{-13} cm$$

While it's experimental radius is

$$r_\alpha = 2.2 \times 10^{-13} cm$$

## 2.2 Non-Machian theories

### 2.2.1 Plasma Cosmology

This cosmology which is often called Alfven-Klein Cosmology, named by two famous physicists Hannes Alfven and Oskar Klein. It tries to explain the entire picture of universe by emphasizing that electrodynamics is as important as gravity for understanding the universe. As it is clear from its name, Plasmas play the main role in this cosmology. Here we give a descriptive explanation of the theory.

### What is Plasma? [6]

An ionized gas which contains negative and positive ions is called plasma. Sometimes it is called as the $4^{th}$ state of matter. The process of ionization takes place after heating the plasma. Plasmas may be completely ionized or partially. When ionization starts, another phenomena named recombination also begins which tries to recombine the ions and finally a state of equilibrium is achieved.

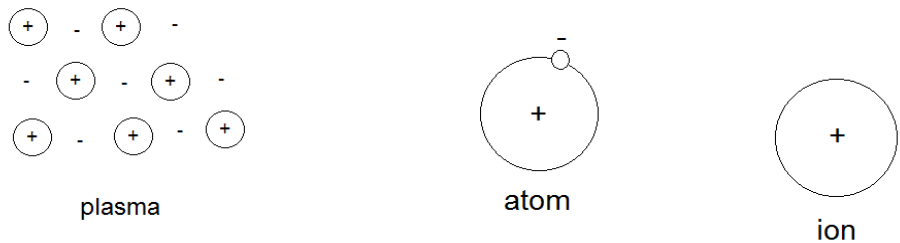

Figure 3-Plasma, atom and ion

It seems that plasmas contain just ordinary matter but we can make plasma by matter and antimatter. Plasma which contains protons and antiprotons (sometimes it also contains electrons and positrons) is called an ambiplasma. One of the important features of this kind of plasmas is that when their particles and antiparticles collide to each other, according to the laws of quantum field theory, annihilation takes place. Ambiplasmas are important parts of plasma cosmology.

## Matter-Antimatter [6]

According to the laws of quantum field theory, there exists an antimatter for every ordinary matter. Antimatters differ from their ordinary ones by their electric charge. Although we have found that antimatter exists and we have succeeded to construct them in our particle accelerators, it is very surprising that we are almost sure that at least in our vicinity in the space there is no antimatter.

SM tries to answer this question by considering that in the very early instants after the Big Bang, there was a very exceed amount of matter. Thus during the annihilation process matter and antimatter annihilated and that very little amount of matter has created our universe. In plasma cosmology the question is answered in another way. Maybe our universe contains both matter and antimatter. At first, this sounds too strange and unphysical but when we think that we even don't know whether stars located in our neighbourhood contain matter or antimatter, this statement becomes reasonable.

But immediately another question comes to mind. If there is some place in our universe that instead of matter contains antimatter, then why it has not been annihilated. The key answer is the Ambiplasma.

## Ambiplasma and matter-antimatter separation [6]

Consider a star and an antistar which is located in it's vicinity. According to the laws of astrophysics, the star emits plasma which contains of ordinary matter. In the same way antistar emits antiplasma contained of antimatter. Now when the plasma and antiplasma meet each other, they collide and an annihilation process begins. The annihilation heats up the plasma in the border region, thus a magnetized ambiplasma is created in that border.

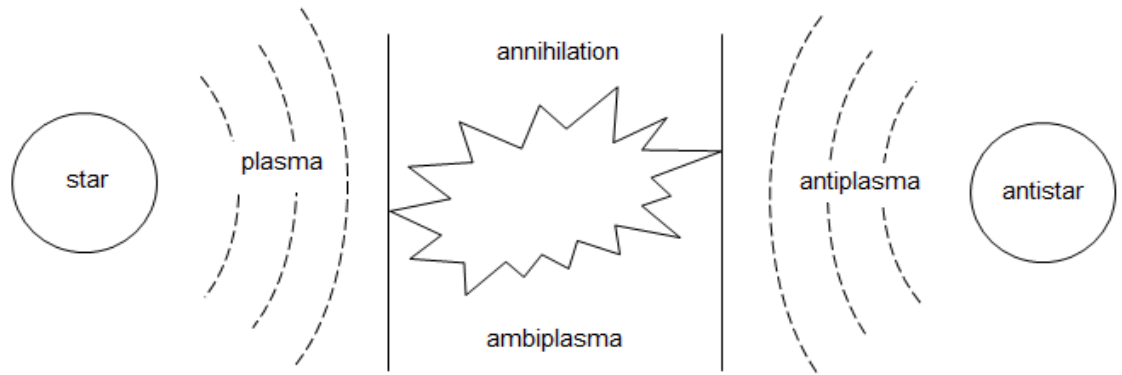

**Figure 4- Separation of star and antistar**

For better understanding what happens after this process, we have to give some remarks on " Leidenfrost layer"

## Leidenfrost layer [6]

Consider a water drop in a plate. It is evident that at temperature T=100°C the drop evaporate making a hissing noise. This evaporation takes place in two different ways.
a- When T grows slowly, the drop becomes smaller and finally it evaporates.
b- When T grows immediately, the drop vanishes in an explosive way.

The difference between these two situations is that in first one, a thin layer is formed which tends to insulate the drop from the plate, this layer is called "Leidenfrost layer"

The same thing happened for collision of plasma and antiplasma, thus a layer formed and separate star from antistar. By this process we can separate the matter from antimatter and be safe of annihilation. Thus it is possible to consider a place in our universe made up of antimatter, which is separated from our part by an ambiplasma.

## Cosmology [5][6]

Unlike SM, which postulates a beginning for our universe, in plasma cosmology there is no beginning and the universe is eternal. Plasma cosmology states that at very long time ago, there was an extremely low density ambiplasma in the universe. Ambiplasma filled in the sphere with a radius of $10^{12}$ light years. Since its density was too law, one proton or antiproton in every $100 m^3$, annihilation was negligible. The only important thing at that time was gravity. Thus all spheres began to contract. When the density became one particle per cubic then there were some chance of collision between particles and antiparticles. Thus little by little the annihilation began. During this annihilation process a huge amount of radiation and heat released. As the radiation was strong enough it could completely cancel the gravitational attraction. So everything was thrown out from the center of attraction. This theoretical approach agrees with Hubble's law. But it explains this effect as a Doppler redshift.

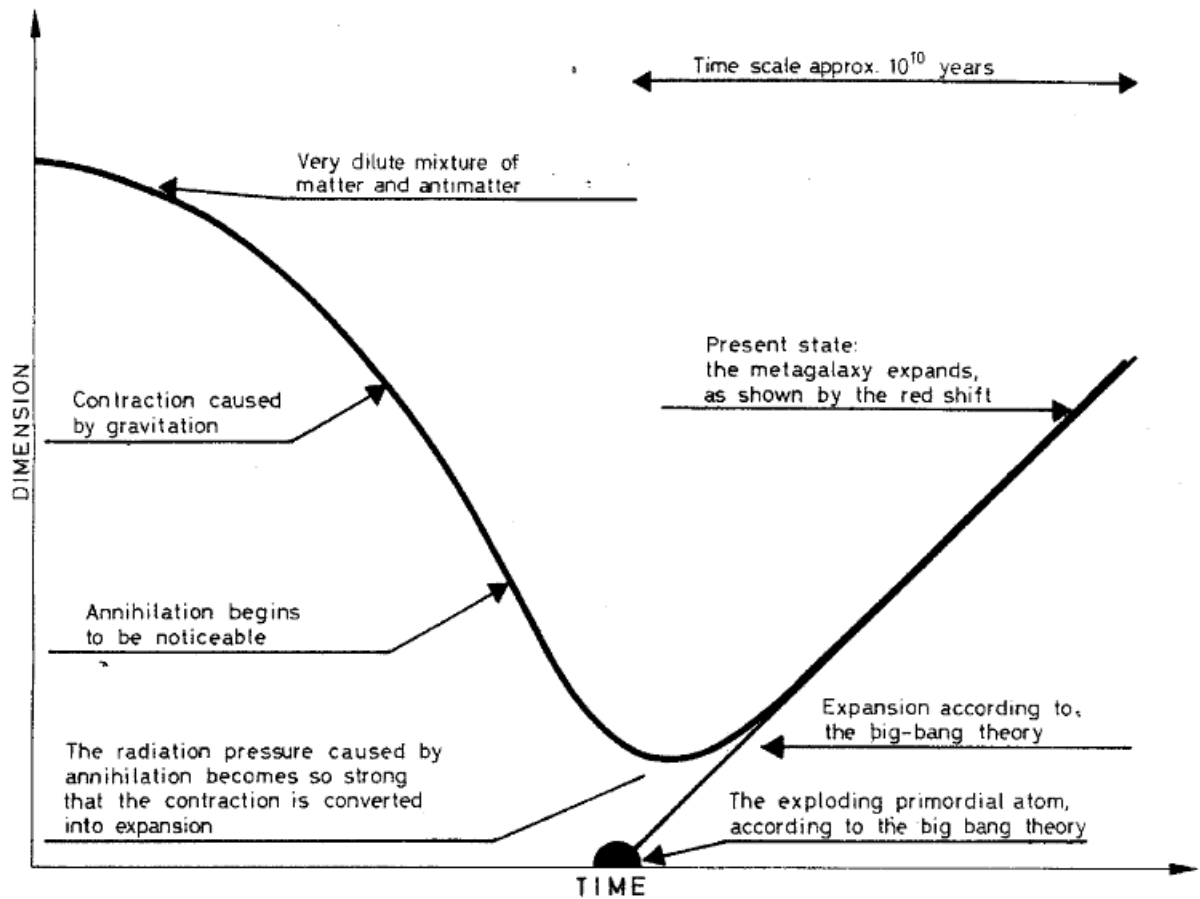

**Figure 5- Development of universe in plasma cosmology and SM**

## Galaxy Formation [6]

It is said that after the annihilation, matter and antimatter dispersed in the universe. Now one can ask that in this situation how galaxies, stars, antigalaxies and antistars can be created. This can be done during a process called 'separation '. Under the action of gravity a layer of heavy ambiplasma is situated at the bottom and the light one goes to the top. An electric current inside the ambiplasma can separate matter and antimatter.

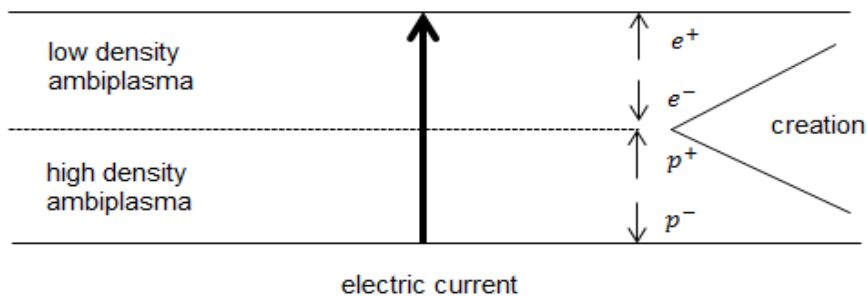

**Figure 6-Matter creation**

Thus we can find a place for creation. It's also remarkable that if we change the direction of electric current, we can find a place for creation of antimatter.

## Finite or Infinite Universe [6]

It has been one of the old questions about the universe that whether the universe is finite or not?

At the beginning of the last century some cosmologists by considering the famous Olbers' paradox about the darkness of the night, concluded that the universe is finite. On the other side Charlier had suggested that universe is composed of systems of increasing dimensions i.e. stars make galaxies, galaxies make clusters, clusters make super clusters and etc. Thus he had concluded that universe is infinite. Plasma cosmology tries to reconcile the infiniteness of universe by Olbers' paradox, simply by stating that universe is infinite and the total mass is infinite too. So its mean density is too low, very close to zero. Most of light we receive is from our local systems i.e. from stars in our galaxy.

## Some observational tests [7]

While the SM does not agree with observations, Big Bang theorists have entered some new concepts like dark matter and dark energy. Dark matter is an object that has no electromagnetic radiation but it is influenced by gravity. They are the missing mass of the matter in the universe. Dark energy is another theoretical object which is responsible for accelerating expansion of the universe. By introducing these concepts, SM cosmologists hope to be able to answer the open problems of cosmology. According to theoretical astrophysics, galaxies organized into filaments and walls that surround large voids. These voids are typically have diameters around 140-170 Mpc and occur with same regularities. Nearly all matter would have to be out of the voids. It has been calculated that if **T** is the time necessary for formation of a void, **R** be its radius, **n** the density of the matter and **V** be the present day velocity, we will have

$$T = 1.03 n^{\frac{-1}{4}} V^{-\frac{1}{2}} R^{\frac{1}{2}}$$

With V=220 km/h, R=85 Mpc and $n = 2.4 \times 10^{-7}/cm^3$ we find T to be 158 billion years, which is 11.6 times larger than the Big Bang. Even if we increase the density by taking to account the amount of dark matter we find T=100 billion years. Same thing happens for dark energy too. In this case if we try to calculate the size of objects, we will find that their size is 35 Mpc. If we try to create voids with sizes typically larger than 60 Mpc their velocity become unacceptable.

## Dark matter [7]

The main reason that in SM's context we believe on dark matter is the constant velocity of objects in spiral galaxies. Thus for this missing part of mass Big Bang theorists entered the concept the dark matter in to context of SM. We have two types of dark matter, cold dark matter (CDM) and hot dark matter (HDM). But until now there has not been detected none of them. Some cosmologists suggested that dark matter is made up of massive neutrinos. But even with taking into account the neutrinos' masses, it will be around 1/10 of the total mass.

Plasma cosmologists associate the missing mass to two dim objects. First is the white dwarf whose mass is in some case $10^{11}$ times massive than solar system and the other is warm plasma that its mass is sufficient for entire local group.

## 2.2.2 Chronometric Cosmology

For this part we almost completely have used the Aubert Daigneault's article.

This cosmology is founded by American mathematician Irving Ezra Segal. Segal's approach was very strange but meanwhile very beautiful. He constructed his theory by some axioms. Although this theory has been refused by large numbers of astronomers and astrophysicists, chronometric cosmology (CC) is remained and regarded as one of the important and elegant theories of modern cosmology.

**Axiomatic Space-time**

Axiom 1: Space-time is a differentiable manifold of dimension 4.

Axiom 2: Space-time is endowed with a causal structure.

For this axiom to be clear, first we have to define some mathematical objects.

- Ordered field : A field ( $F, +, \times$) with a total order $\leq$ is called ordered field if
  i-    If $a \leq b$ then $a + c \leq b + c$ ; $\forall a, b, c \in F$
  ii-   If $0 \leq a$ and $0 \leq b$ then $0 \leq a \times b$

- Convex cone: The set C of a linear vector space, V, over an ordered field, which is closed under linear combinations with positive coefficients is called a convex cone.

The cone C is nontrivial if C and –C have only 0 in common.

This axiom states that on space-time we have a distinguished non trivial convex cone.

Also we can denote p=q which means p temporarily precedes q, if there exists an oriented curve going from p to q.

Axiom 3: Time does not wind back itself.

This means that in our space-time we have no closed timelike loops.

Axiom 4: Causal spatial isotropy

Segal, contrary to all physicists who described the isotropy and homogeneity in a physical and descriptive way, tries to explain them in mathematical language.

For every two arbitrary directions from point P in space-time, not in C(P) nor in –C(P) , there exists a causal diffeomorphism of space-time on to itself that maps one of these directions on to another.

Axiom 5: Causal temporal isotropy

There is no preferred timelike direction at any given point of space-time.

Axiom 6: Space-time can be globally factorized into space and time.

Mathematically this states that there is a diffeomorphism $\varphi$ such that

$$\varphi: R \times S \to M$$

For every $x \in S$, $\varphi$ maps t on $\varphi(x,t)$ which is a timelike arc and for every $t \in R$, $\varphi$ maps x on $\varphi(x,t)$ which is spacelike.

Axiom 7: causal temporal homogeneity

The main goal of this axiom, which as mentioned by A.Levichev very controversial is to state that time translation with respect to factorization of space-time as space× time makes up a causal diffeomorphism.

**Models**

It can be shown that among the famous models, there are only two models which are compatible with these axioms. One is the ordinary Minkowski space, $M_0$, and the other the Einstein's static universe, M.

**Redshift**

CC was refuted by many astronomers and cosmologists because they insisted that neither Minkowskian nor Einstein's universe are not able to explain the redshift. But Segal showed that it is possible to have redshift in CC. Segal found another formula which was different from Hubble's law. For deriving the redshift in CC, first we have to pay attention to both Minkowskian universe and Einstein's one.

It is important to notice that, first one can causally embed $M_0$ into M by a relativistic stereographic projection. And second $M_0$ is tangent to M.

Now it becomes possible to interpret the redshift. In $M_0$, we have

$$M_0 = R \times R^3 \qquad\qquad dS^2 = dx_0^2 - dx_1^2 - dx_2^2 - dx_3^2$$

while in M:

$$M = R \times S^3 \qquad\qquad dS^2 = c^2 dt^2 - r^2 ds^2$$

Where $S^3$ is a 3 dimensional sphere with fixed radius **r**. Now it is evident that the concept of time is different in two models. Calculations show that these two times are related to each other by $x_0 = 2r \tan \frac{t}{2r}$ where **c**=1 and **r** is the radius of sphere. So one deduce that the whole local time line in $M_0$ is a finite interval from $-\pi r$ to $\pi r$ in M.

It is very hard to say that which time we measure in our local observations because their difference is too small.

In CC redshift is defined as

$$\frac{\partial x_0}{\partial t} = 1 + \tan^2 \frac{t}{2r} = 1 + z$$

Although this new redshift is completely different from Hubble's law, Segal showed that the results of his CC theory are compatible with observations.

# 3 Conclusion

At the end of this paper, we want to mention that the above introduced models and theories are not fully compatible with observations, like the standard model. They have some contradictions and unsatisfactory predictions and outcomes.

We would like to insist again that we do not claim that these theories are correct. The existence of them just indicates that we can look to the universe from very different points of view. These different approaches lead us to new theories that will be able to describe some phenomena in our universe like redshift. This may sounds a little strange for some people. This is natural because we have not discovered most regions of our universe; furthermore, we have not been able to test our theories completely (for example we do not know that the General Relativity is valid in the presence of strong gravitational fields or not).

Another important feature is that our method of studying the universe as a whole is very different from other areas of science. In other sciences we do some experiments and observations in the laboratories. Our role is just the observation. We are the observers of the experiment (we do not influence the experiment). But in cosmology the situation is different because we, as an observer, are part of the system which we want to do some investigations i.e. the universe. Also in other branches of physics when people want to do some experiments and achieve an almost precise conclusion, they consider an ensemble for the system. This method is based on the principles of statistical mechanics. But in cosmology it is impossible to make copies of the universe. We only have one.

Considering these facts some scientists doubt about the cosmology as a science. Maybe that is why W.de Sitter has claimed "It should not be forgotten that all this talk about the universe involves a tremendous extrapolation, which is a very dangerous operation" and even perhaps that is the reason that a great authority like L. Landau has said "Cosmologists are often in error, but never in doubt".
Our main goal, as said in introduction, was to give only short and descriptive information about alternative theories to SM.
Often the cosmological models are divided to static and dynamic models, but we decided to divide them into Machian and non-Machian ones because we believe that the Mach's principle has had great influence on physics.
The above mentioned cosmological theories are small part of works of some great physicists like O. Klein, F. Hoyle, J. V. Narlikar, H. Bondi, T. Gold, G. Burbidge, H. Alfven, L. Nottale, I. Segal, E. Lerner, Aubert Daigneault, A. Levichev and many others.
Interested person for further readings may refer to the works, papers and books mentioned in our references.